\documentclass[sigconf]{acmart}
\AtBeginDocument{%
  }

\copyrightyear{2025}
\acmYear{2025}
\setcopyright{rightsretained}
\acmConference[CHI EA '25]{Extended Abstracts of the CHI Conference on Human Factors in Computing Systems}{April 26-May 1, 2025}{Yokohama, Japan}
\acmBooktitle{Extended Abstracts of the CHI Conference on Human Factors in Computing Systems (CHI EA '25), April 26-May 1, 2025, Yokohama, Japan}
\acmDOI{10.1145/3706599.3720038}
\acmISBN{979-8-4007-1395-8/2025/04}

\usepackage{acronym}

\newacro{DHH}[DHH]{d/Deaf and hard of hearing}
\newacro{ASR}[ASR]{Automatic Speech Recognition}
\newacro{WER}[WER]{Word Error Rate}
\newacro{AI}[AI]{artificial intelligence}
\newacro{E2E}[E2E]{end-to-end}
\newacro{HCI}[HCI]{Human–computer interaction}
\newacro{OOD}[OOD]{Out-of-Domain}
\newacro{DNN}[DNN]{deep neural network}
\newacro{HMM}[HMM]{hidden Markov model}
\newacro{RNN}[RNN]{recurrent neural network}
\newacro{ROC}[ROC]{Receiver Operating Characteristic}
\newacro{AUC}[AUC]{area under the curve}
\newacro{ML}[ML]{machine learning}
\newacro{UEQ-S}[UEQ-S]{User Experience Questionnaire}
\newacro{SOTA}[SOTA]{state-of-the-art}

\begin{document}

%%
%% The "title" command has an optional parameter,
%% allowing the author to define a "short title" to be used in page headers.
\title[ASR Confidence Scores for Automated Error Detection]{Evaluating ASR Confidence Scores for Automated Error Detection in User-Assisted Correction Interfaces}

%%
%% The "author" command and its associated commands are used to define
%% the authors and their affiliations.
%% Of note is the shared affiliation of the first two authors, and the
%% "authornote" and "authornotemark" commands
%% used to denote shared contribution to the research.
\author{Korbinian Kuhn}
\email{kuhnko@hdm-stuttgart.de}
\orcid{0009-0005-1296-4987}
\affiliation{%
  \institution{Stuttgart Media University}
  \city{Stuttgart}
  \country{Germany}
}
%\affiliation{%
%  \institution{University of Tübingen}
%  \city{Tübingen}
%  \country{Germany}
%}

\author{Verena Kersken}
\email{kersken@hdm-stuttgart.de}
\orcid{0009-0000-5007-6327}
\affiliation{%
  \institution{Stuttgart Media University}
  \city{Stuttgart}
  \country{Germany}
}

\author{Gottfried Zimmermann}
\email{gzimmermann@acm.org}
\orcid{0000-0002-3129-1897}
\affiliation{%
  \institution{Stuttgart Media University}
  \city{Stuttgart}
  \country{Germany}
}

%%
%% By default, the full list of authors will be used in the page
%% headers. Often, this list is too long, and will overlap
%% other information printed in the page headers. This command allows
%% the author to define a more concise list
%% of authors' names for this purpose.
\renewcommand{\shortauthors}{Kuhn, Kersken, Zimmermann}

%%
%% The abstract is a short summary of the work to be presented in the
%% article.
\begin{abstract}
Despite advances in Automatic Speech Recognition (ASR), transcription errors persist and require manual correction. Confidence scores, which indicate the certainty of ASR results, could assist users in identifying and correcting errors. This study evaluates the reliability of confidence scores for error detection through a comprehensive analysis of end-to-end ASR models and a user study with 36 participants. The results show that while confidence scores correlate with transcription accuracy, their error detection performance is limited. Classifiers frequently miss errors or generate many false positives, undermining their practical utility. Confidence-based error detection neither improved correction efficiency nor was perceived as helpful by participants. These findings highlight the limitations of confidence scores and the need for more sophisticated approaches to improve user interaction and explainability of ASR results.
\end{abstract}


%%
%% Keywords. The author(s) should pick words that accurately describe
%% the work being presented. Separate the keywords with commas.
%% A "teaser" image appears between the author and affiliation
%% information and the body of the document, and typically spans the
%% page.

%% \received{20 February 2007}
%% \received[revised]{12 March 2009}
%% \received[accepted]{5 June 2009}

%%
%% This command processes the author and affiliation and title
%% information and builds the first part of the formatted document.
\maketitle

\section{Introduction}

Confidence scores are an additional output of an \ac{AI} system and represent the degree of certainty a model has in its prediction. In \ac{ASR}, these scores reflect the probability with which a spoken word is recognised correctly. Despite recent improvements in the accuracy of \ac{ASR} \cite{Baevski2020, Radford2023}, transcripts often require manual editing to be intelligible \cite{Butler2019, McDonnell2021, Seita2022}. Automatically highlighting transcription errors could help users to correct them more quickly and accurately. One approach is to use confidence scores to highlight low-confidence words to indicate potential recognition errors \cite{Luz2008A, Luz2008B, Berke2017}. Such automated error detection could help users to correct errors more quickly and accurately. 

\ac{ASR} models can provide word-level confidence scores, which are usually represented as numerical values between 0 and 1, with higher scores indicating greater confidence in the correctness of the result. To automatically classify words as correct or incorrect based on the confidence score, a threshold must be set that defines the minimum confidence level at which a word should be considered correct (e.g. all words above a score of 0.5). As this classification does not necessarily align with the actual correctness of a word, it can lead to false positives and missed detections. Ideally, the threshold is well balanced, assisting users to find major errors without distracting them with non- or minor errors.

Confidence scores are mainly used for downstream \ac{ML} tasks, but have also been proposed to assist users in correcting transcriptions \cite{Luz2008A, Luz2008B} and \ac{DHH} individuals in reading automated captions \cite{Berke2017, Kafle2019}. However, research on their reliability and effectiveness for these tasks remains limited. Evidence from early studies suggests that they are either unreliable \cite{Suhm2001, Feng2004, Vertanen2008} or have a limited positive effect on user performance \cite{Burke2006}. As the architecture of \ac{ASR} has changed fundamentally in recent years, findings from earlier evaluations may not be directly applicable to today's systems. Transformers and \acp{RNN} using attention mechanisms have largely replaced \ac{HMM}-based models in modern \ac{ASR} \cite{Chorowski2015, Chan2016, Kim2017, Gulati2020, Peng2022}, leveraging large-scale deep learning techniques and \ac{E2E} training \cite{Rasmus2015, Baevski2020, Yu2020A}. These \ac{E2E} models show increased robustness and accuracy \cite{Radford2023} but use differing confidence estimation techniques \cite{Park2020} and exhibit overconfidence \cite{Guo2017}. Although the use of confidence scores in \ac{E2E} models for \ac{ML} is an active area of research \cite{DeVries2018, Woodward2020, Kumar2020, Li2021, Tu2022}, their application in user interfaces may differ and has not yet been evaluated.

This study explores whether confidence scores can reliably detect transcription errors and assist users in correcting transcripts. We comprehensively analyse current \ac{ASR} models employing \ac{E2E} architectures. We complement the quantitative evaluation with a user study applying the confidence-based error classification for manual correction. Our results show:

\begin{itemize}
    \item Classifiers using confidence scores achieve only moderate performance and cannot reliably detect transcription errors.
    \item Confidence-based error detection neither improved correction efficiency nor proved useful to participants.
\end{itemize}

\section{Background}

To contextualise previous \ac{HCI} research on confidence scores, it is important to understand the development of \ac{ASR} models and corresponding confidence measures. This section first looks at the technical fundamentals of confidence scores and then at their applied use in \ac{HCI} studies.

\subsection{Confidence Measures in Machine Learning}

Confidence measures are used for various \ac{ASR} downstream tasks, particularly \ac{ML} methods like unsupervised \cite{Yu2020A}, semi-supervised \cite{Chan2004, Huang2013, Park2020}, and active learning \cite{Yu2020B, Huang2016, Li2006}. They are also used for speaker adaptation \cite{DelAgua2018, Uebel2001}, keyword spotting \cite{Benayed2003, Keshet2009, Seigel2013}, and dialogue systems \cite{Hazen2001, Tur2005}. In \ac{HMM}-based \ac{ASR}, confidence is often predicted using binary classifiers on decoding features such as word posterior probabilities \cite{Evermann2000}, word trellis stability \cite{Sanchis2003}, normalised likelihoods, and language model score \cite{Willet1998}. These features are classified using simple linear mappings, such as decision trees \cite{Evermann2000} and Gaussian mixture classifiers \cite{Chigier1992}, or sequence models such as conditional random fields \cite{Seigel2011}, neural networks \cite{Weintraub1997, Schaaf1997}, and recurrent neural networks \cite{Kalgaonkar2015, Li2019, Kastanos2020}.

\ac{ASR} systems have evolved from \ac{HMM} and hybrid designs to \ac{E2E} models, which rely on an encoder-decoder architecture \cite{Chan2016, Kim2017, Dong2018}. A common confidence measure for these models, the softmax probability \cite{Park2020}, is often unreliable due to the overconfident behaviour of \ac{E2E} models \cite{Guo2017, Hendrycks2017, Li2021}. To improve the reliability of confidence scores, a model can be calibrated using temperature scaling \cite{Guo2017}. However, mitigating high confidence predictions for \ac{OOD} data remains challenging and may be an inherent limitation of neural network architectures \cite{Hein2019}. While approaches such as dropout, ensembles, or Bayesian methods can improve uncertainty estimation for some tasks, the accuracy also degrades under dataset shifts \cite{Ovadia2019, Lakshminarayanan2017}. Alternatively, a separate neural network can estimate confidence \cite{Woodward2020, Li2021}, using features from the encoder, decoder, and attention blocks \cite{Kumar2020, Shu2024}. It has also been explored to improve confidence scores around word deletions \cite{Qiu2021, Seigel2014}.

\ac{E2E} models have become the industry standard due to their improved accuracy, but they also exhibit overconfidence. Several techniques can improve their confidence estimation, but none has proven superior. Research suggests that confidence measures become less accurate as \ac{ASR} accuracy decreases. Consequently, transcripts that contain more errors and require more corrections, tend to have less reliable confidence scores.

\subsection{Confidence Scores in HCI}

Despite improvements in \ac{ASR} accuracy \cite{Baevski2020, Radford2023}, transcripts often require manual correction to ensure accessibility \cite{Butler2019, McDonnell2021, Seita2022}. Automatically highlighting errors could speed up editing and improve accuracy. Confidence scores can indicate misrecognised words, but as continuous data, they require a threshold to classify words as correct or incorrect. Choosing an appropriate threshold is a trade-off between precision and recall, aiming to minimise false positives and negatives while maximising true positives.

Early \ac{ASR} systems have already shown a strong correlation between transcription accuracy and confidence scores \cite{Gillick1997}, making confidence-based error detection a long-standing idea \cite{Chase1997}. However, early empirical evaluations suggest that confidence scores are limited in detecting errors and highlight the challenges of determining an effective threshold \cite{Suhm2001, Feng2004}. While highlighting low-confidence words could slightly improve correction accuracy, editing speed remained unchanged \cite{Burke2006}. Error markup also influenced which words users corrected, drawing attention to or away from errors \cite{Vertanen2008}.

Alternatively, confidence scores can highlight correctly transcribed words, helping readers focus on keywords while ignoring errors \cite{Berke2017}. While users expressed a strong interest in such a feature, they found it neither useful nor disruptive \cite{Shiver2015}, with some preferring no markup after use \cite{Berke2017B}. Highlighting "correct" words improved readers' confidence and comprehension \cite{Kipffer2015}, though other studies found no significant benefits from colour-coded text \cite{Vemuri2004}. Underlining and font weight were preferred as markup styles \cite{Kafle2018}, and highlighting 5-15\% of words as important was favoured \cite{Kafle2019}.

Confidence scores have been proposed for correction interfaces \cite{Luz2008A, Luz2008B} and as accessibility markup \cite{Berke2017, Kafle2019}, but evidence supporting these ideas remains limited. Previous studies have highlighted the limitations of then-current \ac{HMM}-based confidence score implementations \cite{Suhm2001}, called for further research \cite{Vemuri2004, Shiver2015, Berke2017B}, and hypothesised that improved confidence accuracy could enable practical applications \cite{Burke2006}. This study contributes to the scarce literature that examines \ac{E2E} models and their modified confidence measures from an \ac{HCI} perspective. We provide a comprehensive evaluation of \ac{SOTA} \ac{ASR} systems and the results of an applied user study.

\section{Methodology}

We first transcribed a public dataset with current \ac{ASR} models that provide word-level confidence scores to evaluate their accuracy for error classification. We then conducted a user study to evaluate the error detection method in practice, its impact on correction efficiency, and its more subjective usefulness. Transcription results, analysis code, and editing prototype are available open-source.\footnote{https://doi.org/10.5281/zenodo.14923041}

\subsection{Dataset}
\label{sec:dataset}

We selected a dataset containing 90 English and 30 German recordings, each three minutes long, with an average of 466 words per sample \cite{Kuhn2024}. The English split was used for a large-scale evaluation and the German split for a user study. We consider the dataset as in-domain data, with realistic audio conditions, resulting in average transcription accuracy and confidence scores by \ac{ASR} systems.

The dataset was transcribed using nine different \ac{ASR} models: Amazon, AssemblyAI, Deepgram, Google, IBM, RevAI, Speechmatics, SpeechText.AI, and Whisper. To the best of our knowledge, all providers have adopted \ac{E2E} models in recent years. All settings were unified to offer the highest accuracy, e.g. selecting large or enhanced models.

\subsection{Analysis}

We used the Levenshtein distance algorithm, which is the basis of the \ac{WER}, to align \ac{ASR} transcripts to a reference solution at the word level. The alignment results are word pairs with an operation (ok, substitution, insertion and deletion), a reference word (except for insertions) and a hypothesis word with confidence values (except for deletions) \cite{Levenshtein1966}. For binary classification, "ok" denotes correctness and other operations indicate errors. The 810 \ac{ASR} transcripts yielded 394,348 word pairs, including 7,489 deletions (which have no confidence score and were excluded from the evaluation). We applied JiWER\footnote{https://github.com/jitsi/jiwer} text transformations to address non-semantic differences, converting text to lowercase, removing punctuation, and expanding contractions. Although Levenshtein-based classification can mislabel correct words due to nearby errors, we expect these errors to have a very limited effect, given the size of the dataset.

\subsection{User Study}

\subsubsection{Design Rationale}

The user study evaluates the effectiveness of confidence-based error markup for transcript correction while minimising confounding variables. A single interface, incorporating findings from prior research, was used. It featured three settings with varying levels of error highlighting: no highlighting, balanced highlighting (balancing correct detections and false positives), and high-threshold highlighting (increasing detections but with more false positives). We used a repeated measures design, where each participant worked with all three interface variants in a randomised order using a Latin square design. Participants corrected transcripts in their native language, German, to avoid biases from second-language use. The task involved real-time editing without audio replay, requiring participants to correct the transcript in one go. This approach was chosen to control for individual effort, preventing it from dominating the results and obscuring the effect of confidence-based error highlighting.

\subsubsection{Participants}

36 people, including undergraduates, postgraduates, and researchers from different universities, participated in the study. All were native German speakers, gave their informed consent, and confirmed they had no hearing, visual, reading, or writing impairments.

\subsubsection{Transcripts}

The transcripts were generated using Whisper, showing high transcription accuracy and average performance for error classification in our evaluation (see Section \ref{sec:results}). Whisper is widely used in scientific research and ensures more comparable and reproducible results. We selected three transcripts from the German dataset split (IDs 70, 82, and 89) with an average accuracy (5.5-6.3\% WER) and a balanced distribution of substitution, insertion, and deletion errors. Balanced thresholds (0.70-0.88) highlighted about 4\% of the words, while a high threshold of 0.99 highlighted 13-21\% of the words (cf. \cite{Berke2017B, Kafle2019}).

\subsubsection{Interface}

\begin{figure}[t!]
  \centering
  \frame{\includegraphics[width=\linewidth]{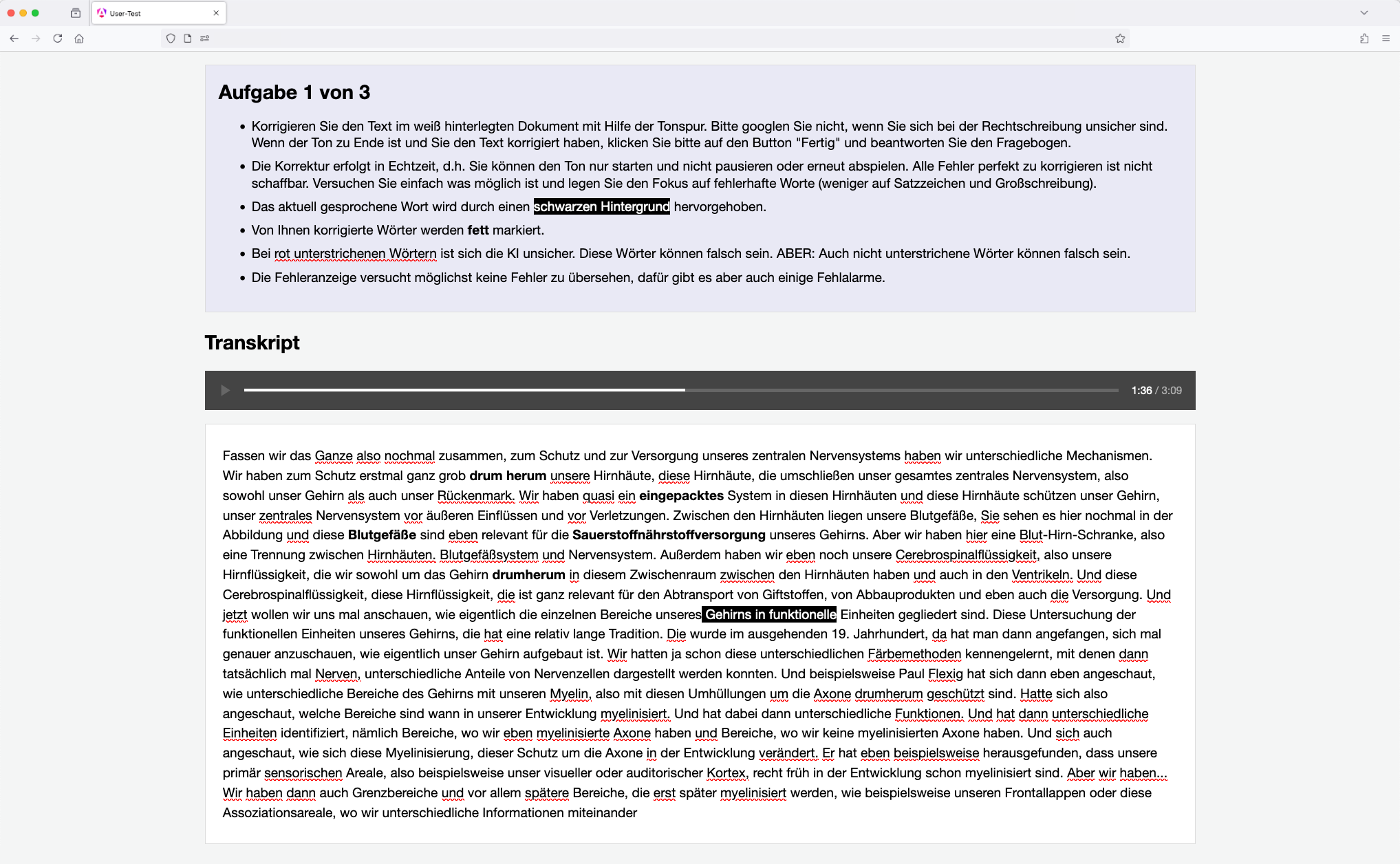}}
  \caption{Correction interface of the user study.}
  \Description{A screenshot of the correction interface used in the user study. The interface is divided into two main sections. At the top is a blue box with bullet points explaining the task and interface features. Below this is an audio player to start the audio recording. Further down, a text editor displays the transcript, with some words underlined with a red wavy line to indicate errors, and others in bold to indicate changes made by the user. Three words in the middle of the text have a black background to indicate that they are currently being spoken in the audio.}
  \label{fig:interface}
\end{figure}

The user interface (see Figure \ref{fig:interface}) featured three markup types: inverted text to indicate the currently spoken words, a wavy red underline for potential transcription errors (similar to spellcheckers), and bold text for user-modified words. Markup styles were informed by prior research \cite{Berke2017B, Kafle2019}, accessibility considerations (e.g., high contrast, colour independence), and familiarity with text editing software while avoiding distractions like text jitter. The text editor was built using Tiptap\footnote{https://github.com/ueberdosis/tiptap} with a custom schema to manage confidence scores, timestamps, and correctness for each word. Edits extended this metadata to track modifications, creations, and deletions, enabling classification of user corrections. User behaviour, such as typing speed and input events, was recorded with timestamps.

\subsubsection{Questionnaire}

After each task, participants completed a questionnaire with 7-point Likert scales. They rated transcription quality and answered four pragmatic quality questions from the short version of the \ac{UEQ-S} \cite{Schrepp2017}, focusing on the efficiency of the interface rather than hedonic aspects. For interfaces with error highlighting, additional scales assessed the frequency, correctness, and usefulness of highlighted errors. An open comment field allowed detailed feedback. After the final task, participants reviewed all questionnaires, adjusted their ratings if needed, and answered which interface option they preferred.

\subsubsection{Procedure}

Participants accessed the prototype via a website containing instructions, interfaces, and questionnaires. They were informed about the study's goals, provided consent, and confirmed that they had no hearing, visual, reading or writing impairments. A technical check ensured audio functionality. Each participant completed three correction tasks, with a fixed transcript order and randomised interface order. Before each task, users received a brief introduction, including explanations regarding the error highlighting functionality (e.g., that the threshold tries to balance or to cover most errors). Participants were informed that highlighted words were not always incorrect and unhighlighted words were not always correct. Editing began with the start of the audio playback, followed by a questionnaire after each task. The study concluded with a final questionnaire and participant debriefing.

\section{Results}
\label{sec:results}

\begin{table*}[ht]
  \caption{Transcription accuracy, confidence correlation, and classification metrics across ASR models.}
  \label{tab:results}
  \centering
  \begin{tabular}{lrrrrrrrrrrrr}
    \toprule
    Model & WER$\downarrow$ & Confidence & $r_s\downarrow$ & $r_{pb}\uparrow$ & Threshold & SD$\downarrow$ & Precision$\uparrow$ & Recall$\uparrow$ & F1$\uparrow$ & AUC$\uparrow$ \\
    \midrule
    Amazon & 0.08 & 0.98 & -0.64 & 0.413 & 0.92 & 0.09 & 0.52 & 0.56 & 0.52 & \textbf{0.87} \\
    AssemblyAI & 0.06 & 0.92 & -0.42 & 0.274 & 0.55 & 0.16 & 0.46 & 0.36 & 0.33 & 0.78 \\
    Deepgram & 0.12 & 0.94 & -0.84 & \textbf{0.539} & 0.81 & 0.10 & \textbf{0.55} & 0.59 & \textbf{0.55} & 0.86 \\
    Google & 0.24 & 0.94 & -0.57 & 0.314 & 0.94 & \textbf{0.05} & 0.41 & 0.50 & 0.44 & 0.68 \\
    IBM & 0.15 & 0.87 & \textbf{-0.88} & 0.515 & 0.77 & \textbf{0.05} & 0.52 & 0.60 & 0.54 & 0.84 \\
    Rev AI & 0.08 & 0.95 & -0.73 & 0.490 & 0.87 & 0.09 & 0.49 & \textbf{0.64} & 0.54 & \textbf{0.87} \\
    Speechmatics & \textbf{0.05} & 0.99 & -0.63 & 0.427 & 0.87 & 0.12 & 0.49 & 0.50 & 0.47 & 0.80 \\
    SpeechText.AI & 0.13 & 0.98 & -0.78 & 0.396 & 0.93 & 0.06 & 0.54 & 0.41 & 0.46 & 0.69 \\
    Whisper & 0.08 & 0.94 & -0.61 & 0.505 & 0.70 & 0.22 & 0.48 & 0.54 & 0.47 & 0.83 \\
    \bottomrule
  \end{tabular}
\end{table*}

\subsection{General Analysis}

\subsubsection{Dataset Distribution}

An Anderson-Darling test was conducted to check if the distribution of word-level confidence scores followed a normal distribution. The test result showed a statistic of 74901.069, exceeding the 0.787 critical value at the 5\% significance level. A skewness test (statistic: -458.746, p < .001) showed a significant negative skew, with frequent scores clustered towards the high confidence scores. We conclude that the dataset is not normally distributed and has significant long tail and right skew characteristics. This is as expected, as most words are correctly transcribed, and \ac{E2E} models exhibit overconfidence. Consequently, non-parametric tests were chosen for the general analysis.

\subsubsection{Transcript WER and Confidence Score}

Table \ref{tab:results} shows the mean \ac{WER} and confidence score of all transcripts by \ac{ASR} model. Spearman’s rank correlation was computed to assess the relationship between transcript-level \ac{WER} and confidence score. For all models, \ac{WER} was significantly related to confidence score (see Table \ref{tab:results} column $r_s$, all $ps < 0.001$). The correlation was moderate to strong and varies between the models independently of their mean \ac{WER}. It can be concluded that the mean confidence score is higher for transcripts with a lower \ac{WER}.

\subsubsection{Word Correctness and Confidence Score}

A point-biserial correlation analysis was conducted for each \ac{ASR} model to assess the relationship between confidence score and word correctness. For all models, confidence scores were significantly related to word correctness (see Table \ref{tab:results} column $r_{pb}$, all $ps < 0.001$). The correlation was weak to moderate. The results suggest that words with higher confidence scores are more likely to be correct.

\subsubsection{Classification Performance}

We used the F1-score to evaluate the classification performance of confidence scores in detecting transcription errors. This metric ignores true negatives, which are less relevant in this context, as the classification task focuses on error detection within a dataset that mostly contains correct samples. For each transcript, the threshold that achieved the highest F1-score was used. The optimal thresholds varied between models, and some models also showed considerable variability (see Table \ref{tab:results} columns Threshold, SD). The classification performance of precision, recall, F1-score, and \ac{AUC} are shown in Table \ref{tab:results}. The models achieved precision scores ranging from 0.41 to 0.55, indicating a 50\% chance that detected errors were actual errors. Recall scores ranged from 0.36 to 0.64, suggesting that approximately 50\% of transcription errors were missed. F1-scores ranged from 0.33 to 0.55, reflecting poor to moderate classification performance, indicating that the classifier has poor to moderate performance, balancing precision and recall, but not excelling at either. The \ac{ROC} \ac{AUC} values, ranging from 0.68 to 0.87, indicate a moderate fit of the classifier.

\subsection{User Study Results}

\subsubsection{Word Error Rate Reduction}

Participants' corrections reduced the \ac{WER} on average by 0.47 ($SD=0.16$) without error highlighting, 0.48 ($SD=0.14$) with balanced-threshold highlighting, and 0.45 ($SD=0.17$) with high-threshold highlighting. A repeated-measures ANOVA was performed to evaluate the effect of the interface on the relative \ac{WER} reduction. Mauchly’s test indicated that the assumption of sphericity had been met, $\chi^2 (2) = 1.61, p = .447$. The results show that the \ac{WER} reduction was not significantly affected by the type of interface, $F(2, 70) = 0.27, p = .767, \eta^2 = .005$.

\subsubsection{Modification Ratio of Highlighted Errors}

Chi-squared tests of independence were performed to assess the relationship between highlighted errors and user modifications. For the balanced-threshold interface, highlighted errors were significantly more likely to be modified, $\chi^2(1) = 30.32, p < .001$, with 2.64 times higher odds (CI 95\% [1.87, 3.73]). For the high-threshold interface, no significant association was found, $\chi^2(1) = 1.18, p = 0.278$, with 1.31 times higher odds (CI 95\% [0.84, 2.03]).

\subsubsection{Transcription Quality Rating}

Participants rated transcription quality on a seven-point Likert scale (1 = very bad, 7 = very good). Mean ratings were 5.9 ($SD=1.3$) without highlighting, 5.8 ($SD=1.0$) with balanced-threshold highlighting, and 5.4 ($SD=1.3$) with high-threshold highlighting. A repeated-measures ANOVA was performed to evaluate the effect of the interface on the perceived transcription quality. Mauchly’s test indicated that the assumption of sphericity had been met, $\chi^2 (2) = 4.0, p = .136$. The results show that the perceived transcription quality was not significantly affected by the type of interface, $F(2, 70) = 1.71, p = .188, \eta^2 = .028$.

\subsubsection{Interface Efficiency Rating}

Participants rated pragmatic quality using four seven-point Likert scales from the \ac{UEQ-S}. Mean scores were 1.47 ($SD=1.16$) without highlighting, 1.51 ($SD=1.17$) with balanced-threshold highlighting, and 1.41 ($SD=1.08$) with high-threshold highlighting. A repeated-measures ANOVA was performed to evaluate the effect of the interface on the \ac{UEQ-S} pragmatic quality. Mauchly’s test indicated that the assumption of sphericity had been violated, $\chi^2 (2) = 9.97, p = .007$, therefore Greenhouse-Geisser corrected tests are reported ($\epsilon = .80$). The results show that the UEQ pragmatic quality was not significantly affected by the type of interface, $F(2, 70) = 0.21, p = .765, \eta^2 = .001$.

\subsubsection{Usefulness of Error Highlighting}

Participants rated error highlighting on three seven-point Likert scales: frequency (1 = far too few, 7 = far too many), correctness (1 = very wrong, 7 = very correct), and usefulness (1 = very irritating, 7 = very helpful). Paired t-tests showed significantly better ratings for the balanced threshold compared to the high threshold condition in frequency ($M = 4.2$ vs. $M = 5.3$, $t(35) = -4.84, p < .001$), correctness ($M = 3.6$ vs. $M = 2.9$, $t(35) = 2.58, p = .014$), and usefulness ($M = 4.1$ vs. $M = 3.3$, $t(35) = 2.60, p = .014$).

\subsubsection{Preferred Interface Type}

After completing all tasks, participants rated their preferred interface. Two participants had no preference. 16 participants preferred no error highlighting, 18 balanced-threshold highlighting, and 4 high-threshold highlighting. In the optional comments section, 16 participants provided feedback on the error highlighting feature. Three expressed initial optimism about confidence-based highlighting before using it, 10 found it irritating, and 7 reported ignoring it after some time.

\section{Discussion}

We evaluated \ac{E2E} \ac{ASR} models to assess the reliability of confidence scores in detecting transcription errors. The results showed a strong correlation between \ac{WER} and confidence at the transcript level, with higher confidence scores associated with fewer errors. However, confidence and word correctness showed only a weak to moderate correlation, highlighting the challenge of accurately classifying errors at the word level. 

Setting a confidence threshold to classify word correctness is challenging due to the heavily skewed dataset. Most words have high confidence scores, and the confidence ranges for correct and incorrect words largely overlap. Thus, low thresholds miss most errors, while high thresholds produce many false positives. The inability to find a balanced threshold has already been reported for \ac{HMM}-based \ac{ASR}, and our results indicate that \ac{E2E} models show similar limitations (cf. \cite{Feng2004}).

We calculated the most effective threshold for each transcript based on the performance of the F1-score. The average threshold varied between and within the \ac{ASR} models. This variance will likely increase in real-world scenarios, as our evaluation was limited to a single dataset. In practice, error classifiers would likely use suboptimal thresholds, as the optimal threshold is unknown during transcript correction. For the best thresholds, precision, recall, and F1-scores hovered around 0.5, indicating moderate success in identifying transcription errors. Precision shows that only half of the detections were actually errors, while recall shows that half of the errors were missed, highlighting the challenge of achieving high accuracy and coverage. \ac{ROC} \ac{AUC} values of 0.68 to 0.87 also indicate moderate classification performance, comparable to an \ac{HMM}-based \ac{ASR} evaluation study \cite{Suhm2001}. The low observed accuracy could lead users to distrust the system \cite{Yin2019}, and the presence of misses makes the classification unpredictable, a crucial aspect for the explainability of \ac{AI} \cite{Waa2020}.

The user study showed that participants corrected about 50\% of transcription errors, regardless of the error highlighting setting. Correction accuracy did not differ significantly between settings, suggesting that visual cues did not affect their ability to detect and correct errors. Highlighting effectively drew attention to words, as long as highlighting was not too frequent (cf. \cite{Vertanen2008}). Participants generally rated the transcription quality as high, and these ratings were not significantly affected by the interface configuration. Thus, error highlighting, regardless of its frequency or accuracy, did not influence overall confidence in transcription accuracy. 

The \ac{UEQ-S} results showed good pragmatic quality with no significant differences between settings, suggesting that the interface was perceived as efficient regardless of the presence or absence of error highlighting. In the balanced threshold condition, users rated the frequency, correctness and usefulness of highlighted errors positively. In the high threshold condition, highlighting was perceived as excessive, largely incorrect and irritating, indicating a preference for a more moderate approach. Most participants preferred either no highlighting or the balanced threshold condition. Similar to previous studies on keyword highlighting for \ac{DHH} individuals \cite{Shiver2015, Berke2017B}, participants found the concept of confidence-based error highlighting promising at first, but not very useful or even distracting after experiencing it.

\section{Conclusion}

Confidence scores have been proposed to automate error detection and assist users in correcting transcriptions. This study evaluated the reliability of these scores for error classification through a comprehensive analysis of current E2E ASR models and a user study examining their application in practice. 

The results showed that confidence scores correlate with word correctness but are unreliable to guide manual correction. The overlap in confidence scores between correct and incorrect words, combined with the low proportion of incorrect words, makes it difficult to set a threshold and limits classification performance. Classifiers showed moderate success at best, failing to detect most errors or producing many false positives. The user study confirmed the general analysis, as confidence-based highlighting neither improved correction efficiency nor was perceived as helpful.

We conclude that current \ac{ASR} confidence scores are insufficient for automated error detection and can be misleading, especially for those without a background in \ac{ML}. This work aims to help \ac{HCI} researchers understand the technical background, interpretation, and limitations of these scores in explainable \ac{AI}.

\section{Limitations \& Future Work}

The rapidly evolving field of \ac{ASR} may lead to models and confidence measures that better reflect transcription errors. However, while \ac{E2E} models have significantly improved \ac{ASR} accuracy, our evaluation showed that confidence scores remain unreliable, suggesting a more fundamental problem. The results are limited to a single dataset, focusing on English and German under more favourable audio conditions, limiting generalisability to other languages. However, more extreme scenarios are likely to degrade classification performance \cite{Ovadia2019}. Although the user study examined a real-time editing scenario, the classifier also appears ineffective for non-live correction tasks where users iteratively correct transcripts \cite{Luz2008A}. 

\ac{ASR} is a valuable accessibility tool, but cannot signal uncertainty like human interpreters. Future research should focus not only on accuracy, but also on how \ac{ASR} can communicate errors to increase the explainability of the results.

%%
%% The acknowledgments section is defined using the "acks" environment
%% (and NOT an unnumbered section). This ensures the proper
%% identification of the section in the article metadata, and the
%% consistent spelling of the heading.
\begin{acks}
We thank all participants for their time and contribution to this research. We thank Roland Mangold for his valuable thoughts on the statistical analysis. This research was conducted as part of the SHUFFLE Project and funded by "Stiftung Innovation in der Hochschullehre".
\end{acks}

%%
%% The next two lines define the bibliography style to be used, and
%% the bibliography file.
\bibliographystyle{ACM-Reference-Format}
\bibliography{bibliography}

%%
%% If your work has an appendix, this is the place to put it.
% \appendix

\end{document}